\documentclass[conference]{IEEEtran}
\IEEEoverridecommandlockouts
\usepackage{authblk}
\usepackage[noadjust]{cite}
\usepackage{amsmath,amssymb,amsfonts}
\usepackage{algorithmic}
\usepackage{graphicx}
\usepackage{textcomp}
\usepackage{float}
\usepackage{xcolor}
\usepackage{listings}
\usepackage{multirow}
\usepackage{graphicx}
\usepackage{amssymb}
\def\BibTeX{{\rm B\kern-.05em{\sc i\kern-.025em b}\kern-.08em
    T\kern-.1667em\lower.7ex\hbox{E}\kern-.125emX}}
\begin{document}

\title{A Conceptual Reference Model for Human as a Service Provider in Cyber Physical Systems\\
}

\author[1,2]{Hargyo T.N. Ignatius}
\author[1]{Rami Bahsoon}
\affil[1]{\textit{School of Computer Science}, \textit{University of Birmingham}, United Kingdom \authorcr Email: {\tt \{hxi903, r.bahsoon\}@cs.bham.ac.uk}\vspace{1.5ex}}
\affil[2]{Dept. of Computer Engineering, Universitas Multimedia Nusantara, Indonesia}


\maketitle

\begin{abstract}
In Cyber Physical Systems humans are often kept in the loop as operators and/or service users. Yet in many cases, humans and machines collaborate and provide services to each other. Research on service models and service composition for CPS exist; however, humans as service providers have not been adequately considered as part of the CPS service composition model. We provide a classification of human-as-a-service in CPS, and we propose a Service Oriented Architecture (SOA) ontology model for the CPS environment as part of the Everything-as-a-Service paradigm. The model considers human characteristics and their dynamics, as a service provider or collaborator with the machine. As the ontology model is an enabler for engineering a self-adaptive CPS with human-machine collaboration as service providers, we describe how a commonly used self-adaptive reference model can be refined to benefit from the vision. We evaluate the ontological contribution against criteria that relates to accuracy, completeness, adaptability, clarity, and consistency. We demonstrate the feasibility of our conceptual reference model using a use case from the medical domain and we show how human-machine service provision is possible. 
\end{abstract}


\section{Introduction}
Cyber physical system (CPS) is one of enabling technologies for Industry 4.0 along with Artificial Intelligence, Cloud, and Big Data. Among several definitions, CPS can be described as ``engineered systems that are built from, and depend upon, the seamless integration of computation and physical components" \cite{nsf2020cps}. Engineering self-adaptivity in CPS has to consider various sources of uncertainties in dynamic environments \cite{nunes2015survey, musil2017patterns}, within or across more than one CPS layer \cite{Muccini2016May}.

In smart manufacturing, as a case, CPS often involves humans; for example, as part of the machine-human feedback loop(s). Human-machine collaboration provides flexibility that allows manufacturers to adapt more easily to shifting demands in products and processes \cite{wef2020hilcps}. Though fully automated CPS can excel in strength, precision and speed, humans with cognitive abilities, consciousness, and skills can adapt more quickly to new requirements and tasks. 

Humans and machines differ in many aspects. Humans work based on their consciousness, while machines operate based on what is taught/programmed. Humans have to work based on a motive which is often the result of a trade-off analysis between rewards and risks. Humans have free will so that humans can decide to stop working or choose to do work differently based on the context and their considerations. Besides, many factors influence human performance, such as mood, fatigue, incentives, etc.

In the many common CPS use cases, humans, when kept in the loop, are generally an operator; the users who instruct or initiate requests for and receive services from CPSs (\textit{service consumers}). However, many complex CPS is essentially a combination of computers, machines, and people who work together to achieve system goals \cite{Sowe2016cphs}.  
In such systems, human can provide services by performing tasks based on their ability to sense, act, store, and process data (e.g. citizen sensing, citizen actuation \cite{crowley2013hlcps}). Therefore, humans in the loop can be viewed as not only the service consumers but also as the \textit{service providers} \cite{zhou2020cpss}. The increasing interest and use cases coming from various disciplines have made Human-in-the-loop a branch of research in CPS, widely known as Human-in-the-Loop CPS (HiTLCPS) \cite{nunes2018practical}. Cyber-Physical-Social Systems (CPSS) \cite{zhou2020cpss} is among the subareas of HiTLCPS.

Service Oriented Architecture (SOA) provide potential solutions for modelling, run time synthesis, management and composition of HiTLCPS to deal with variability in multitude types of component types and changing application environments at runtime \cite{wang2020soa}. With SOA, every capability possessed by each entity is considered as either atomic or composite service. However, traditional SOA models and composition technique have limitations when directly applied to CPS for various reasons due to the heterogeneity of physical entities, whether human or machine, while considering context requirements, service provision constraints, and services similarity.

Several works have proposed service model and service composition for CPS \cite{Huang2009soa, Wang2014soa, feljan2015soa, Zhu2015soa} and CPSS \cite{wang2020soa}. However, these studies do not pay much attention to human as a service provider in CPS. Human characteristics are not explicitly modelled. Humans are mostly considered part of the physical entities, along with robots, vehicles, sensors, and other actuators. Indeed, existing models are not adequate to accommodate humans as service providers, and new or enhanced models are needed.

The novel contributions of this paper are as follows: We first define the problem of human As a Service in CPS service composition and we motivate its need.  We contribute to a novel reference service-oriented Service Oriented Architecture (SOA) ontology model for the CPS environment as part of the Everything-as-a-Service paradigm. The model considers human characteristics and their dynamics, as a service provider or collaborator with the machine. The model builds on existing service composition paradigms and extends it beyond the machine-centric ones to also include human-as-services in CPS. As the ontology model is an enabler and pre-requisite for engineering a self-adaptive CPS with human-machine collaboration as service providers, we describe how a commonly used self-adaptive reference model, MAPE-K can be refined to realise the vision. The proposal is a pragmatic shift towards acknowledging that both human and machines work in collaboration as service providers.  The paradigm can enable new modalities of services composition, where human can assist the machine (vice-versa is also true), considering some qualitative attributes such accumulated experience, knowledge, skill, abilities, and other human attributes such as emotion, mood, compassionate, fatigue, etc.

The ultimate vision is to transit the problem of service composition into a collaborative human-machine service composition, where bidirectional infosymbiotic cooperation/learning between the machine and human can be envisioned, promising more dependable and human-centric CPS services provision. We report on how the model can be instantiated using a use case from the medical domain. We follow the standard and commonly used approaches to ontology evaluation, where we evaluate the ontology against criteria that relates to accuracy, completeness, adaptability, clarity, and consistency.    


\section{*-as-a-Service in HiTLCPS  }
\subsection{Everything as a Service}
Everything as a service (XaaS) is a concept for services and applications that users can access over the network, which is generally found in the form of Software-as-a-Service (SaaS), Platform-as-a-Service (PaaS), and Infrastructure-as-a-Service (IaaS). However, in its development, we also see more specific terms such as Communication-as-a-Service (CaaS) which provides VoIP services, Transportation-as-a-Service (TaaS) such as online taxi or ride-hailing services, and many others.

HiTLCPS integrates computation, networking, and physical processes that involve human in the loop. Mobile internet devices with varying computing capabilities are strong candidates for implementation in physical entities of the CPS \cite{la2010soa}. Some machines may have limited computing resources, but many devices could execute complex computation and processes. These devices can communicate and share services over the network. The development of ubiquitous computing technology allows human-computer interaction to a higher level with various interfaces. Humans can be accessed and interacted with the system through handheld devices or other human interface devices (HID) nearby. A service-oriented architecture is therefore promising to HiTLCPS to enable collaboration between components in providing services.

\subsection{Human-as-a-Services}
The idea of human-as-a-service is supporting the XaaS (Everything-as-a-Service) paradigm that sees that humans can provide services to the system; so can other devices. 

Human-as-a-service in CPS is defined as \textit{a ``thing" of Everything as a Service with human capabilities and properties}. These are humans as service providers that can work either in isolation or in collaboration with machines in CPS to sense, process computation, actuate, learn and/or transfer its learning with the objective of providing more socio-dependable and human-centric service composition models for CPS. The relation can be collaborative or an arms race, based on the context with the incentive of a better overall service provision. 

Human-as-a-service is widely manifested as an individual or group of services, often by direct appointment or through an open-call (crowdsourcing) mechanism. It exhibits unique characteristics as it evolves during its life cycle and involves various ways of collaboration/communication \cite{huang2016human, huang2018human}.

Human-as-a-services within the CPS can vary. Treating CPS as a self-adaptive system using the MAPE-K \cite{ibm2006mapek} reference architecture, human-as-a-services can be applied within all layers of monitoring, analyze, planning, execution, and knowledge. Nunes et al. \cite{nunes2015survey} have identified several human roles in the loop that we use as a reference to categorize human-as-a-services in HiTLCPS as follows:

\begin{enumerate}
    \item  \textit{Sensor service}. In their activities, humans might use tools and computing devices equipped with digital sensing functions. Humans also have five natural senses that can detect many events (e.g. traffic hours, car accidents, fires, etc.). Humans can provide this service actively by reporting an event or phenomenon detected by the five senses and passively by allowing their activity/behaviour to be recorded to see social phenomena (social sensors).

    \item  \textit{Processing service} - Humans are learning creatures who have developed cognitive abilities. With his diverse knowledge and intuition, human choices will help make decisions, especially when dealing with uncertainties due to lack of knowledge or other environmental dynamics.

    \item  \textit{Actuating service} - In everyday life, humans already act as actuators. When receiving an emergency signal from the patient room, the nurse will immediately go to the patient room and take the necessary actions. Within the scope of HiTLCPS, sensor networks or robots may detect errors and require specialized actuation from humans to fix the problem \cite{nunes2015survey}.
    
    \item \textit{Adaptation Service} - This service is a composite of the three services above. Humans can act as an adaptation promoter for other nodes.  ``The users (with different roles) may decide whether the adaptation is needed, which strategy to choose, and even participate in its realization" \cite{kazhamiakin2010adaptation}. This role includes but not limited to, control feedback, provision of knowledge, learning, and evaluation.
\end{enumerate}

\section{The O*NET Framework}

Understanding the classification and relationship between the attributes of workers and their jobs is essential to build an adequate human-as-a-service model and pre-requisite for developing a self-adaptive model. We have studied several existing frameworks, namely O*NET \cite{onet2020matching}, SOC \cite{soc2020matching}, and ISCO \cite{isco2020matching}. We have decided to use O*NET as it considered to be the primary reference for building human capability models.

\begin{figure}
\centering
\includegraphics[width=0.49\textwidth]{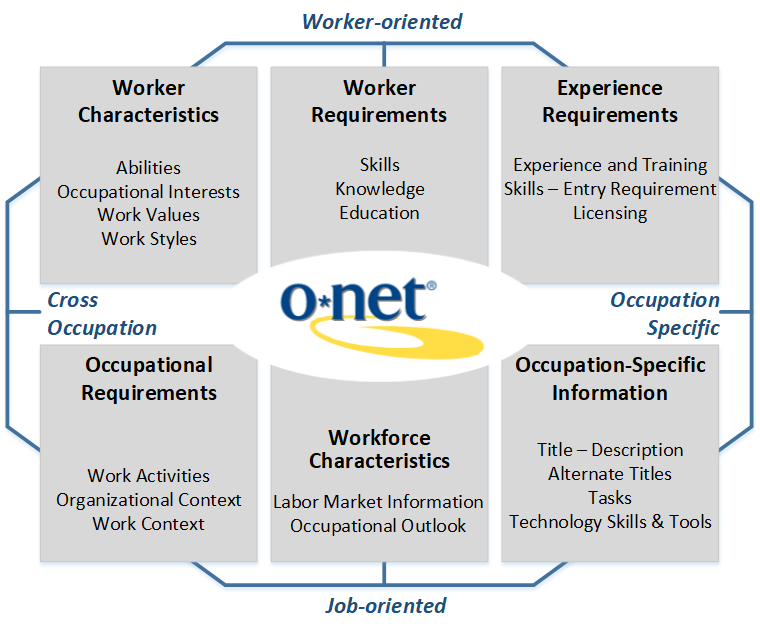}
\caption{The O*NET content model \cite{onet2020matching}} \label{fig:onet_contentmodel}
\end{figure}

The Occupational Information Network (O*NET)  provides a rich set database of occupation information that describes the job and worker characteristics. The Content Model defines the most important types of job information and incorporates them into a theoretical and empirical framework.

In O*NET framework, every single job requires a different selection of knowledge, skills, and abilities and performs a variety of tasks and activities. These particular characteristics of an occupation are described by the Content Model (as seen on figure \ref{fig:onet_contentmodel}, which reflect the characters of occupations (job-oriented descriptors) and people (via worker-oriented descriptors).

Worker-oriented descriptors consist of several attributes as follows:
\begin{enumerate}
    \item Worker Characteristics are defined as enduring features that can affect both performance and the capacity to learn the knowledge and skills necessary for the efficient performance of the job. These characteristics are classified as follows:
    \begin{enumerate}
        \item Abilities: enduring attributes of the person that affect performance.
        \item Occupational Interests: Preferences for conditions/environments at work.
        \item Work Values: Global aspects of work consist of basic needs that are essential to an individual's satisfaction.
        \item Work Styles: Personal features that can influence how well someone does a work.
    \end{enumerate}
    \item Worker Requirements reflect an individual's developed or acquired qualities that may be correlated with work performance.These attributes are categorized as follows:
    \begin{enumerate}
        \item Skills: Developed capacities that promote learning (faster acquisition of knowledge) and performance of activities that occur across jobs.
        \item Knowledge: organized sets of concepts and facts applying in general domains.
        \item Education: Prior academic experience needed to perform in a job.
    \end{enumerate}
    \item Worker Experience Requirements are previous work experiences that involve employee experiential backgrounds such as certification, licensing, and training.
    \begin{enumerate}
        \item Experience and Training: relevant work experience, apprenticeship, and on-site/on-the-job training required.
        \item Skills-Entry Requirement: entry requirement for developed capacities that facilitate learning and performance.
        \item Licensing: awarded licenses, certificates, or registrations to show that a job holder has acquired certain skills.
    \end{enumerate}
\end{enumerate}

\section{Proposed SOA-HiTLCPS Ontology Model}

We view HiTLCPS as a combination of humans and machines who interact, communicate, and collaborate to complete their tasks. We use the term machine to refer to any computing system with networking capabilities designed to meet its task cycle autonomously. The machine can be cloud systems and smart devices that are close together in a work environment. To create a self-adaptive human-machine service provision, we need to have a pre-requisite model that includes both human and machine capabilities. To simplify semantic discovery and reasoning, we propose an SOA model for human-in-the-loop CPS, which is expressed as an ontology, called the SOA-HiTLCPS ontology model.

\begin{figure}
\centering
\includegraphics[width=0.485\textwidth]{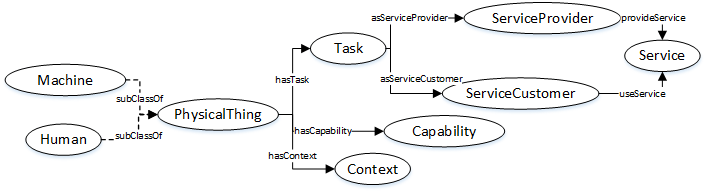}
\caption{Human-machine relationship in HiTLCPS} \label{fig:human_machine_interaction}
\end{figure}

Figure \ref{fig:human_machine_interaction} is a top ontology of our proposed SOA-HiTLCPS ontology model which explains that humans and machines are within an organization where each node has its function and task which generally correspond to its context. \textit{Tasks} are roles and activities that have goals to be achieved. In carrying out their roles and duties, each \textit{Physical Thing} may provide services (act as Service Providers) or use services (Customer Service). \textit{Capabilities} are things that enable humans/machines to complete their tasks well. \textit{Context} is the environment, background, setting, or surroundings of events or occurrences of the tasks. Context can be a physical location, time, temperature, and other contexts in a broader scope related to tasks.

During the process of achieving its goals, the human/machine may need services from others. For example, a bomb disposal technician needs robotic services to cut cables. Or vice versa the robot needs the services of the bomb squad to decide which cable to cut. We can see that each node can be a \textit{Service Provider} or a \textit{Service Customer}.

\subsection{Service Model}
\begin{figure}
\centering
\includegraphics[width=0.4\textwidth]{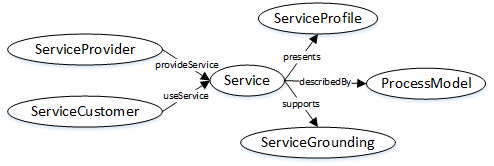}
\caption{Upper layer service ontology} \label{fig:upper_layer_model}
\end{figure}

For interoperability reasons, we propose a service model following the OWL-S, which consists of three main parts, namely \textit{Service Profile}, \textit{Process Model}, and \textit{Service Grounding} as in figure \ref{fig:upper_layer_model}. 

The \textit{Service Profile} describes what the service does, and the parameters used, such as input, output, preconditions, effects, service limitations, and non-functional characteristics that distinguish it from other similar services.  

The \textit{Process Model} is a specification that explains how the service is used, what constraints must be satisfied and what patterns are required to interact with the service. 

\textit{Service Grounding} describes how to interact with the service (message format, transport protocol, etc.). In OWL-S, the service grounding is a bridge between syntax- and protocol-oriented WSDL and semantics-oriented OWL.

\subsection{Service Profile}
\begin{figure}
\centering
\includegraphics[width=0.485\textwidth]{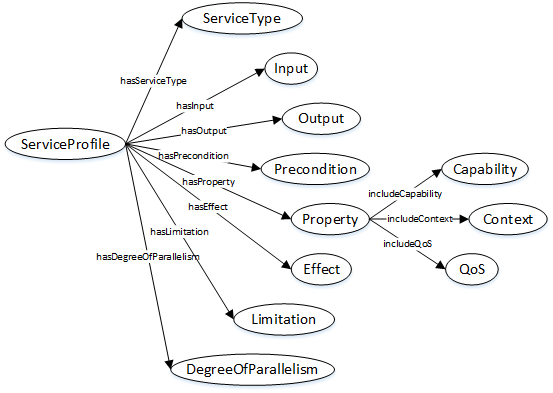}
\caption{Service profile} \label{fig:service_profile}
\end{figure}

Service Profile allows providers to advertise their services, and also requester to specify the service capabilities they require. The aim is to support the Service Discovery mechanism to find the most suitable service-customer needs. Each element in Service Profile in figure \ref{fig:service_profile} is described as follows:
\begin{itemize}
    \item \textit{Service Type} describes the types of service that can be either atomic or composite. Atomic service can be in the form of sensing service, actuating service, or communicating service. Meanwhile, composite service is a combination of several atomic services.
    \item \textit{Input} are the data that the service requires as the input to process with.
    \item \textit{Output} are the data produced by the service.
    \item \textit{Preconditions} are all conditions that must be met (true) before service execution.
    \item \textit{Property} is an attribute that is held by the actor/service-provider at that time. These attributes can be related to context, capability model (discussed in the next subsection), and QoS (e.g. reputation, cost, response time, etc.)
    \item \textit{Effects} are conditions that hold after the service execution.
    \item \textit{Degree of Parallelism}, borrowed from \cite{Sun2016soa}, indicates the number of requests this service can serve.
    \item \textit{Limitations} are things that limit the continuity and availability of services. For example, a service can only be delivered within a specific time frame, a certain distance, particular location and condition
\end{itemize}

\subsection{Human Capability}
We argue that human-as-a-service is closely related to occupation because, in essence, humans provide services in every task they do within the scope of their profession.

An occupation could involve one or more human-as-a-services, atomic and composite. As the human qualification determines the quality of work, we consider it necessary to put knowledge, abilities and skills as essential components in our proposed human capability model.

\begin{itemize}
    \item \textit{Characteristic} represents psychophysiological factors \cite{klinestiver1980psychophysiological} that distinguish humans and affect the services provided. 
    \\
  
    We express these factors into three categories:
    \begin{itemize}
        \item \textit{Preferences} correspond to a person's preferences for work environments and outcome that could affect service availability such as time, location, price. Preferences are compatible with O*NET's \textit{occupational interets}. 
        
        
        \item \textit{Abilities} express innate human attributes that affect their cognitive, physical, psychomotor, and sensory performance. These ability attributes are usually defined with a measurement scale. Abilities are compatible with O*NET's \textit{abilities}.
        
        \item \textit{Performance factors} are internal and external variable, aspects of human behaviour and the context (or environment), that can affect human performance reliability. This element is a derivative of the \textit{Work Value} and \textit{Work Style} in the O * NET framework. Scale is used to describe which factors are more dominant than others. 
    
    \end{itemize}
    \item \textit{Qualification} are attributes that describe a person's appropriateness/fitness, achievement, and quality, which can be either Skill sets, Knowledge, formal Education, or Experience.
    \begin{itemize}
        \item \textit{Skills} are obtained from training and experience which are defined together with the scale.
        \item \textit{Knowledge} refers to domains of expertise or scope/area of work. This pair of qualifications is essential. As an illustration, someone with driving skills and knowledge of city A will find it difficult to drive in city B.
        \item \textit{Education} refers to one's level of formal education or degree.
        \item \textit{Experience} stands for records of services that have been performed along with ratings obtained from service requesters. The rating system used can vary and may include several assessment criteria.Referring to the O * NET framework, Experience also records practical training (i.e. on-site/on-the-job training).
    \end{itemize}
    \item \textit{Potential} is defined as is defined as latent human capacity to improve, for growth and developement \cite{bates2001redefining}. Human skills are developed by knowledge acquired from experience. Not only improving the quality of services in general (improve the skills level), this also opens up new service opportunities (\textit{Potential Service}) that may be provided after new knowledge and skill are acquired. This concept is alligned with the concept of Maximum Human in \cite{paleri2018active} to maximize the 
    active humans for greater returns in their activity profile, also with Human Capability Theory \cite{vogt2005maximizing} in which social systems should promote human flourishing.

\end{itemize}

\begin{figure*}
\centering
\includegraphics[width=0.8\textwidth]{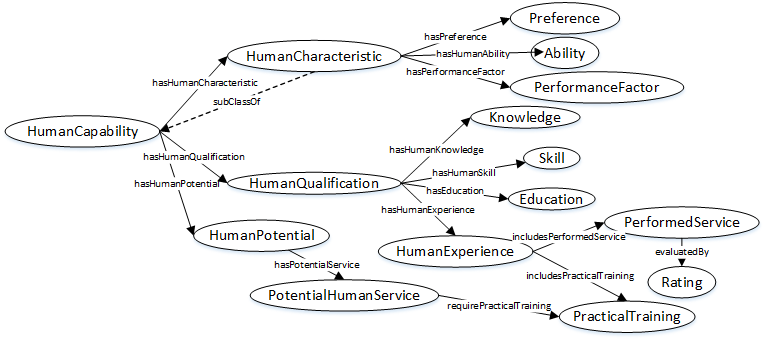}
\caption{Human capability model} \label{fig:human_capability}
\end{figure*}

\subsection{Machine Capability}
Its hardware and software specifications define the computing capability of a machine. Analogously this is similar to Abilities in humans, but the machine can be upgraded with better component replacement.
Skills and knowledge on the machine are the programming logic and datasets provided by the creator. If AI technology is employed, then machines can grow their knowledge (i.e. dataset, ontology) to improve their ability to perform certain functions/services. Machine learning can be done online or offline using shared artefacts or inferred during communication with other nodes. 
However, to acquire a new type of skill, new logic needs to be inserted into the system. In other words, without reprogramming the machine will not have new services automatically.

\section{Using our Ontology in Self-adaptive HiTLCPS}
The definition of our SOA-HiTLCPS ontology model is a pre-requisite for supporting future developments for a self-adaptive human-machine service provisioning in CPS. Self-adaptivity in HiTLCPS can relate to bi-directional cooperation in which machines can help humans or vice versa. Therefore, it is essential to understand machine vs human behaviour to properly utilize their strengths in a collaborative-oriented environment for optimal results (i.e., not a competition to replace each other).

We instantiated the model using a simple scenario in the context of smart health care CPS environment. The CPS system connects patients, medical experts, and other smart agents (i.e. machines).  We implemented the proposed model as a semantic information model by leveraging OWL standard ontology language and Protégé \cite{protege2020dec} editor to evaluate the feasibility of our conceptual model. For space limitations, we do not provide instances of all concepts; only those are essential to demonstrate the feasibility and applicability of our model.

\subsection{Architecture}
\begin{figure*}
\centering
\includegraphics[width=0.8\textwidth]{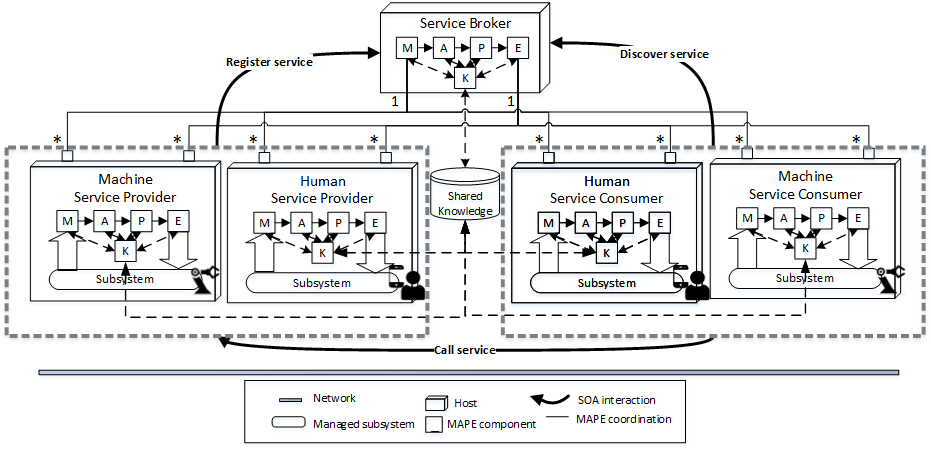}
\caption{Generic SOA model for self-adaptive systems using MAPE-K hierarchical control pattern}
\label{fig:soa_sas}
\end{figure*}
Depending on the domain characteristics and requirements, several self-adaptation control-loop patterns can be used \cite{lemos2013}, be it hierarchical control, master/slave, regional planner, information sharing, or fully decentralized.

We show how our ontology model can enrich IBM's MAPE-K reference architecture, where we use the Hierarchical Control pattern. In figure \ref{fig:soa_sas}, two layers of MAPE-K (Monitor-Analyze-Plan-Execute over
a shared Knowledge) feedback loop is implemented using the SOA paradigm. Although it looks like a master/slave pattern, in a hierarchical control pattern, the overall system is controlled by a hierarchical control structure, where each hierarchy level has complete MAPE-K loops.

MAPE-K loops at different levels interact by exchanging information that contributes to new knowledge stored in a shared knowledge base that other hosts can access.

The top layer is a service broker that carries out service provisioning, composition, discovers and invokes service implementation candidates that meet the criteria requested and returns the best invocation result or composition plan to service consumer.

The second layer is self-adaptive systems with MAPE-K loops that interact directly with managed resources or subsystems, representing machines and humans. Depending on the service flow direction, each self-adaptive system at this hierarchical layer can be a Service Provider (when delivering services) and Service Consumer (when using services).

In service provisioning, the second layer coordinates with the adaptive service broker to achieve optimal results. The hierarchical structure allows the second layer to focus on more concrete adaptation goals, while the higher level can handle adaptation strategies for a broader perspective \cite{weyns2013patterns}.

\subsection{Example Scenario}  
A health clinic uses a web chat as a health service channel. Patients can take advantage of this service to ask questions about clinical services, doctor schedules, book a GP, and get health advice.

In this scenario patient $Adam$, doctor $David$, and chatbot $Cathy$ are at layer two that exchange and utilize services from one another. Patients and doctors are represented by smart personal devices that provide an interface (e.g. API) for other entities to interact with their users.

Patient $Adam$ accesses the webchat service to consult about the health problems he is experiencing. For every new conversation session request, the $chatbotService$ by chatbot $Cathy$ is allocated first.

$Cathy$ answers $Adam$'s questions relying on its AI and knowledge from the local and shared knowledge base. During the conversation, $Cathy$ acquires information from the chat with $Adam$ and stores it on a shared knowledge base for others to use.

$Adam$ repines of discomfort in his head, yet $Cathy$'s answers don't quite satisfy him. With natural language processing and through its MAPE-K loops, $Cathy$ detects $Adam$'s emotions and upset. $Cathy$ then sends a \textit{service discovery} request to the \textit{Service Broker} with several criteria to maintain customer experience and satisfaction. Based on the conversation, $Cathy$ can infer that $Adam$ needs a human service with better ``Medicine and Dentistry", ``Therapy and Counseling" knowledge and ``Complex Problem Solving" skills which $Cathy$ does not pose.  

Based on the criteria given by \textit{Service Broker} discovers and invokes service implementation candidates that meet the invocation criteria using the SPARQL query as follows:
\
\begin{lstlisting}[
    basicstyle=\footnotesize, %or \small or \tiny etc.
]
SELECT ?service
WHERE {
    ?service soa-hitlcps:presents ?serviceprofile .
    ?serviceprofile soa-hitlcps:hasProperty ?property .
    ?property soa-hitlcps:includeCapability ?capability .
    ?capability soa-hitlcps:hasHumanSkill ?skill .
    ?capability soa-hitlcps:hasHumanKnowledge ?knowledge 
    FILTER (?skill=soa-hitlcps:Complex_Problem_Solving 
    &&  ?knowledge IN (soa-hitlcps:Medicine_and_Dentistry,
    soa-hitlcps:Therapy_and_Counseling))
}
\end{lstlisting}
The query above returns the result that the $chatDoctor$ service by doctor $David$ is a suitable candidate. The Service Broker invokes the $chatDoctor$ service for $Cathy$, which then gives $David$ access to join the chat session. At this point, $Cathy$ becomes a \textit{Service Consumer} (to $David$) as well as a \textit{Service provider} (to $Adam$).

$David$ provides an \textit{adaptation service} for the chatbot $Cathy$ to provide a better quality of service in the future. $Cathy$ acquires the knowledge from the conversation between patient $Adam$ and doctor $David$. This knowledge can be in the form of questions, answers and responses given. However, during the chat session, $Cathy$ can still provide answer recommendations which can be adjusted by $David$.

$David$ gives some advice to $Adam$, and they agree on a schedule for offline meetings. $Adam$, who was initially upset, ends the session with good satisfaction.

\begin{figure}
\centering
\includegraphics[width=0.48\textwidth]{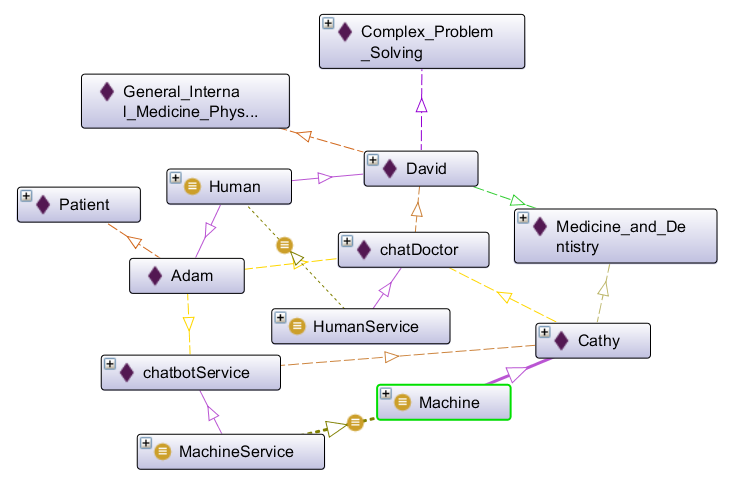}
\caption{Individuals and classes relationships for Scenario 2.}
\label{fig:sec2_instant}
\end{figure}

The instantiation of this scenario is shown in figure \ref{fig:sec2_instant}.

By involving chatbot $Cathy$, doctor $David$ can get other work done, handle more customers, thus increase productivity. $Cathy$ can learn from $David$ to provide more human responses in the future by showing empathy. Meanwhile, $Cathy$ can also offer answer suggestions to $David$, especially when it comes to data processing where humans are inferior to AI. 

This scenario shows that machines and humans can help each other extend the capabilities of both. Self-adaptive HiTLCPS benefiting from our ontology can modify its behaviour or structure in response to the changing context (environment, goal, system) that they perceive.

\section{EVALUATION}



We provide a qualitative and quantitative evaluation of our ontological model referring to \cite{vrandevcic2009ontology} to insure our design adheres to certain desirable criteria; such as accuracy, completeness, adaptability, clarity, and consistency. We use HermiT 1.4.3.456 ontology reasoner, which can evaluate whether or not the ontology of input is consistent, defines subsumption relationships between classes, and much more. Several test scenarios are employed by adding several instances/individuals and the relations that will connect one individual to another. The expected results then be compared with the Inferred Results by HermiT.

\subsubsection{\textbf{Accuracy}} is a criterion that indicates whether the axioms of an ontology comply with the domain knowledge. We have made every effort to make each element or concept expressed in Class or Relations in this model comply to existing standards and literature. For instance, in our human capability model, we refer to O*NET for a taxonomic approach and integrate it with the Human Capability Theory to add the $HumanPotential$ concept.
    
To provide accuracy we ensure that the axioms should constrain an ontology's potential interpretations such that the resulting models are consistent with the users' conceptualizations. In the illustration below, it can be seen that axioms 1,2 refers to the facts and conceptualization of the concept in the given concept definition 1, 2.
\\\\
\textit{
Given:\\
Concept 1: Human is a physical thing with human capability \\
Concept 2: Human service is a service provided by human\\
Output:\\
Axioms 1: PhysicalThing and (hasCapability some HumanCapability) SubClassOf Human\\
Axioms 2: Service and (providedBy some Human) SubClassOf HumanService
}    
    
\subsubsection{\textbf{Completeness}} measures if the domain of interest is appropriately covered. The domain of interest of this model is to promote human-machine service provisioning so that we ensure our ontology is able to answer several basic competency questions (CQ) related to service delivery. These competency questions are formulized as SPARQL queries towards ontology. We compare our model with the existing related models, HSCD \cite{Sowe2016cphs} and PE-ontology \cite{Huang2009soa} in table \ref{tab:ontology_comparison}. Our model provides the answers to the given CQs, while the other two models require a change in the ontology and develop the concept into several subclasses of ontology (i.e. ontology evolution) to answer questions related to skills, knowledge, and abilities. HSCD does not provide a model for machines/physical entities because it limits their scope to humans, whereas PE-ontology does not take into account humans specifically in their model.
\begin{table}[h]
\tiny
\caption{Comparison of our SOA-HiTLCPS model with other models in answering Competency Questions}
\resizebox{\columnwidth}{!}{%
\begin{tabular}{|l|l|l|l|}
\hline
\multicolumn{1}{|c|}{\multirow{2}{*}{Competency   Questions}} & \multicolumn{3}{c|}{Provision   of Answer}                                                    \\ \cline{2-4} 
\multicolumn{1}{|c|}{}                                        & \multicolumn{1}{c|}{SOA-HiTLCPS} & \multicolumn{1}{c|}{HSCD} & \multicolumn{1}{c|}{PE-ontology} \\ \hline
CQ1: Is this human?                                           & \checkmark                     & \checkmark                & R/E                              \\ \hline
CQ2: Is this machine?                                         & \checkmark                     & R/E                       & \checkmark                       \\ \hline
CQ3: Which human as this ability?                             & \checkmark                     & R/E                       & R/E                              \\ \hline
CQ4: What services are available?                             & \checkmark                     & \checkmark                & \checkmark                       \\ \hline
CQ5: Which service requires this skill?                       & \checkmark                     & R/E                       & R/E                              \\ \hline
CQ6: Which service requires this knowledge?                   & \checkmark                     & R/E                       & R/E                              \\ \hline
CQ7: Which services meet the given criteria?                  & \checkmark                     & \checkmark                & \checkmark                       \\ \hline
CQ8: Who is the service provider for this service?            & \checkmark                     & \checkmark                & \checkmark                       \\ \hline
\end{tabular}%
}
\newline
\newline
\checkmark : instant, R/E: requires evolution 
\label{tab:ontology_comparison}
\end{table}


\subsubsection{\textbf{Adaptability}} measures how far the ontology anticipates its use. It should offer the conceptual foundation for a range of anticipated tasks and allow for methodologies for extension, integration, and adaptation. New tools and unexpected situations should be able to use the
ontology. Our proposed ontology enables not only human-machine service provisioning but also machine-only service provisioning, human-only service. However, ones can leverage existing concepts in our ontology for other purposes. Our SOA-HiTLCPS's human capability is closely related to human resource development functions. Several concepts can be utilized for better provision of training, career planning, promotion, and payroll. Another example is predictive maintenance that primarily involves foreseeing the system's breakdown to be maintained by detecting early signs of failure to make maintenance work more proactive. Some techniques like oil analysis, vibration analysis (mechanical looseness or weakness) are possible by leveraging the $Machine_Specification$ and $Experience$ concepts on our model. 

\subsubsection{\textbf{Clarity}} measures how effectively the ontology communicates the intended meaning of the defined terms. This criterion can be measured by using Class/Relation Ratio (CRR) from \cite{gangemi2005ontology} that can be formulised as:
\begin{equation}
 CRR(O)=\frac{|C(O)|}{|P(O)|} \label{eq:2}  
\end{equation}
where C(O) is the cardinality of the set of classes represents by nodes in O, and P(O) is the cardinality of the set of relations in O.

We compare our SOA-HiTLCPS model with HSCD, and PE-ontology in table \ref{tab:CRR_comparison}. Each class and relation in the models is assumed as a class and object property in the ontology. Although actually Object Properties, Equivalent Classes, Disjoint Classes, Subclasses (Subclass of) are counted as relationships, for an apples-to-apples comparison we only calculate Object Properties and Subclasses to determine P(O). In this illustration, we can see that our model involves more classes and relations than the other two models with the lowest CRR. Lower CRR value means there are more relations/properties to explain a concept (class); provides more clarity.

\begin{table}[h]
\caption{Class/Relation Ratio (CRR) Comparison}
\tiny
\resizebox{\columnwidth}{!}{%
\begin{tabular}{|l|r|r|r|}
\hline
                    & \multicolumn{1}{c|}{\textbf{SOA-HiTLCPS}} & \multicolumn{1}{c|}{\textbf{HSCD}} & \multicolumn{1}{c|}{\textbf{PE-ontology}} \\ \hline
\textbf{C(O)}       & \textbf{46}                             & \textbf{17}                        & \textbf{10}                               \\ \hline
Object   properties & 45                                      & 10                                 & 9                                         \\ \hline
Subclasses          & 10                                      & 6                                  & 0                                         \\ \hline
\textbf{P(O)}       & \textbf{55}                             & \textbf{16}                        & \textbf{9}                                \\ \hline
\textbf{CRR(O)}        & \textbf{0.84}                           & \textbf{1.06}                      & \textbf{1.11}                             \\ \hline
\end{tabular}%
}
\label{tab:CRR_comparison}
\end{table}

\subsubsection{\textbf{Consistency}} describes that the ontology does not include or allow for any contradictions. We ensure consistency using two different methods. First, we rely on the output of HermiT reasoner \cite{glimm2014hermit}, which is based on ``hyper tableau" calculus that provides efficient reasoning and ontology consistency tests. Moments after the HermiT reasoner is started HermiT will generate errors if it finds any inconsistencies, and our implementation is free of this. We also ensure there are no inconsistencies by providing no class equivalent to owl: Nothing in the inference results.

\begin{figure*}
\centering
\includegraphics[width=0.7\textwidth]{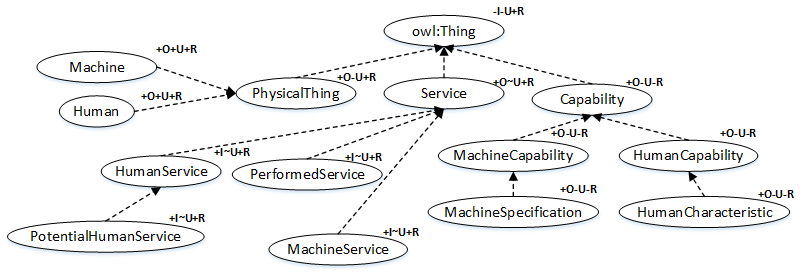}
\caption{The subsumption relationships with their Ontoclean metaproperties}
\label{fig:ontoclean}
\end{figure*}

Second, we use the Ontoclean \cite{guarino2004ontoclean} methodology to analyze the taxonomy of classes that have subsumption relations (i.e. sub-class, sub-type). Ontoclean has the following rules: given two properties, $p$ and $q$, when $q$ subsumes $p$ the following constraints apply:
if $q$ has anti-rigid (\textbf{\textasciitilde R}) and/or anti-unity (\textbf{\textasciitilde U}) and/or an identity (\textbf{+I}) criterion and/or a unity criterion (\textbf{+U}) then $p$ must carry the same corresponding criterion/metaproperty. As shown in figure \ref{fig:ontoclean}, the subsumption relationship in our ontology is consistent, according to Ontoclean.

\section{Related Research}
Huang et al. \cite{Huang2009soa} presented a Physical-Entity service-oriented architecture model to enables inter-operation and coordinated sharing of distributed and heterogeneous data using Physical-Entity (PE) ontology that classifies the physical entities (including human) and their class properties and services.  The proposed Service Model follows the OWL-S model, including the Input, Output, Precondition, and Effect specifications. However, this model divides precondition and effect into context precondition, non-context precondition, non-context effect, and context effect. The non-context effects is a change in the world or environment after service execution. Meanwhile, the context effect is a change in the service provider entity after performing the service. There are service provision constraints that represent the physical constraint of the PE relevant to the service, such as maximum distance, maximum load, etc. However, human characteristics and capabilities that affect human's services quality have not been accommodated in this model.

Echoed \cite{Huang2009soa}, Wang et al. \cite {Wang2011soa} proposed an ontology model for context-sensitive specification of the service abilities of physical entities. Physical Entities (PE) provide atomic services with behaviour constraints such as context precondition,  precondition, postcondition.  Context preconditions correspond to preconditions related to the dynamic context of the PE that should establish before PE can provide the services. Precondition accounts the service constraints irrelevant to the context where the postcondition is the conditions after a service's execution. This model does not adequately accommodate human-as-a-service as humans can provide composite services.

Zhu et al. \cite{Zhu2015soa} extended OWL-S ontology concerning several significant issues related to CPS / IoT where every Physical Thing (PT) entity can provide a service, as well as receive the impacts from any service.  PT entities are described in four main classes, namely ``Physical Profile", ``Operation Profile", ``Operation Schedule", and ``Context". To simplify the reasoning process, Zhu introduced the ``AppliedTo" concept to the Service model. The ``AppliedTo" class represents recipient PT and effects that can change recipient PT's states after the service's execution. The model also does not consider humans as part of PT.

Sun et al. \cite{Sun2016soa} proposed an ontology-based CPS service model with location, physical entities and CPS services as the three main components in which the CPS service uses physical entities and has effects on them. This model pays more attention to the location alongside the state of physical entities as context. Several characteristics for physical entities were introduced, such as operation space (working region), degree of parallelism (whether this physical entity can be used by more than several services simultaneously), working state (the availability state of the physical entity). Contrary to our proposal, the physical entities are part of the CPS service, not the service provider in this model.

A human service capability description (HSCD) model has been proposed by Sowe et al. \cite{Sowe2016cphs}. The main objectives of this model are to represent the person's identity, the tasks a person can perform, the qualifications of the person for performing the tasks, the types of interfaces can be used to interact with the person. In this model person's capability relies on qualification, certification, and rating. In our SOA-HiTLCPS model, we break down the capability into abilities, skill, knowledge, and other characteristics that may affect a person's performance.
\section{Conclusion}
We propose a conceptual SOA ontology model for human as a service provider in Cyber Physical Systems, called SOA-HiTLCPS ontology model. In our model, machines and humans can help each other extend their capabilities; humans can provide sensing, processing, actuating, and promote adaptation for other nodes within the CPS. A use case scenario from the medical domain is used to illustrate how SOA-HiTLCPS can be instantiated. As SOA-HiTLCPS is an enabler and pre-requisite for engineering a self-adaptive CPS with human-machine collaboration as service providers, we have reported on how self-adaptive reference architectures models such as MAPE-K can be refined and leverage SOA-HiTLCPS. In addition to establishing feasibility and applicability of SOA-HiTLCPS by means of instantiation on a use case and enrichment of MAPE-K, the paper follows standard and commonly used approaches to ontology evaluation, where we evaluate the ontology against criteria that relates to accuracy, completeness, adaptability, clarity, and consistency. 

Our ongoing work is looking at further refinements and implementation of the model. Our future research may also incorporate different aspects of QoS and evaluation against existing adaptive systems. The ultimate objective is to provide efficient and dependable self-adaptive human-machine service provisioning. We look at how structured or unstructured human-machine data can be dynamically analyzed and consolidated to drive the adaption in human in the loop CPS. 

\section*{Acknowledgment}
This work was supported by the Indonesia Endowment Fund for Education (Lembaga Pengelola Dana Pendidikan – LPDP). The authors would like to thank the anonymous reviewers whose detailed and insightful comments helped improve and clarify this manuscript.

\bibliographystyle{IEEEtran}
\bibliography{references}
\end{document}